\documentclass[12pt,british,sort&compress]{iopart}
\usepackage[T1]{fontenc}
\usepackage[latin9]{inputenc}
\usepackage[a4paper]{geometry}
\usepackage{color}
\usepackage{babel}
\usepackage{verbatim}
\usepackage{float}
\usepackage{amsbsy}
\usepackage{graphicx}
\usepackage{esint}
\usepackage[numbers]{natbib}
\usepackage[unicode=true,pdfusetitle,
 bookmarks=true,bookmarksnumbered=false,bookmarksopen=false,
 breaklinks=false,pdfborder={0 0 1},backref=false,colorlinks=true]
 {hyperref}

\makeatletter

\providecommand{\tabularnewline}{\\}
\newcommand{\lyxdot}{.}

\makeatother

\begin{document}
\title{The effects of shape and amplitude on the velocity of scrape-off
layer filaments}

\author{JT Omotani$^{1}$, F Militello$^{1}$, L Easy$^{2,1}$ and
NR Walkden$^{1}$}

\address{$^{1}$ CCFE, Culham Science Centre, Abingdon, Oxon, OX14
3DB, UK}

\address{$^{2}$ Department of Physics, University of York, Heslington,
York YO10 5DD, UK}

\ead{john.omotani@ccfe.ac.uk}

\begin{abstract}

A complete model of the dynamics of scrape-off layer filaments will
be rather complex, including temperature evolution, three dimensional
geometry and finite Larmor radius effects. However, the basic mechanism
of $\boldsymbol{E}\times\boldsymbol{B}$ advection due to electrostatic
potential driven by the diamagnetic current can be captured in a much
simpler model; a complete understanding of the physics in the simpler
model will then aid interpretation of more complex simulations, by
allowing the new effects to be disentangled. Here we consider such
a simple model, which assumes cold ions and isothermal electrons and
is reduced to two dimensions. We derive the scaling with width and
amplitude of the velocity of isolated scrape-off layer filaments,
allowing for arbitrary elliptical cross-sections, where previously
only circular cross-sections have been considered analytically. We
also put the scaling with amplitude in a new and more satisfactory
form. The analytical results are extensively validated with two dimensional
simulations and also compared, with reasonable agreement, to three
dimensional simulations having minimal variation parallel to the magnetic
field.

\end{abstract}

%\pacs{52.55.Dy,52.65.Kj,52.55.Rk}

%\noindent{\it Keywords\/}: filament, SOL, tokamak

%\submitto{\PPCF}

%\ioptwocol

\setcounter{footnote}{0}

Filaments are a prominent feature of the scrape-off layer (SOL) of
tokamaks, in L-mode and in H-mode both during and between ELMs\citep{ben-ayed2009},
and in other magnetised plasmas. They provide a significant component
of the particle transport\citep{boedo2003}, especially in the far
SOL, and so may have a strong impact on particle fluxes to the first
wall and divertor. Data from gas-puff imaging diagnostics\citep{myra2013}
suggests that filaments may sometimes have significantly elliptical
(rather than circular) cross-sections. Most of the theoretical work
on SOL filaments has been done in simplified, two dimensional models,
see \citep{krasheninnikov2008} and \citep{dippolito2011} for reviews,
but recently there has been an increasing amount of attention given
to more realistic models, for example using three dimensional simulations\citep{angus2012,walkden2013,easy2014}
or including finite Larmor radius effects\citep{madsen2011}. These
developments have motivated us to reexamine and extend the earlier
analytical work to try to give as complete a physical picture of the
two dimensional mechanisms that regulate filament motion as possible,
in order to facilitate the interpretation of more complicated models
applied to isolated filaments, and ultimately to SOL turbulence simulations\citep{ricci2013,militello2013,tamain2015}.

Our subject here is the scaling of filament velocity with various
parameters of the filament, in the two dimensional limit where parallel
variation may be neglected and assuming cold ions and isothermal electrons.
This represents the very simplest model that can capture the basic
mechanism of filament motion. By characterising quantitatively the
basic physical processes driving filament motion in this simple model,
we will be able, when analysing more complicated models, to identify
the deviations from this behaviour. These deviations can then be ascribed
to the extra physics in, for example, three dimensional or non-isothermal
simulations. The aim therefore is to provide a theoretical tool to
aid the interpretation of more complicated models, as part of a programme
of research building systematically towards models which are both
well enough understood to be trusted and also realistic enough to
be quantitatively compared to experiment and used predictively for
future machines. The particular model we consider contains two mechanisms
that limit the filament velocity: inertia and sheath currents. While
inertia is universally present, the sheath current (as modelled here)
is only relevant on the assumption that parallel resistivity is small
so that parallel currents may reach the sheath unimpeded. Where this
assumption is invalid, whether due to cold plasma\citep{easy2015}
or interaction with neutrals in the divertor or to large magnetic
shear near X-points, different mechanisms will come into play; for
some examples see \citep{krasheninnikov2008}. If it is desired to
include other effects, as different closures for the two dimensional
equations, we hope that the framework of the calculation provided
here will make further additions relatively straightforward.

We present here a calculation of the scaling of the maximum filament
velocity allowing arbitrary elliptical cross-sections, which have
not previously been considered analytically. We also aim to clarify
the physical mechanisms while giving more detail than has previously
been published. In particular, the scaling with amplitude, though
considered in \citep{kube2011} and mentioned briefly in several works\citep{garcia2005,myra2005,theiler2009},
has not previously been given a satisfactory analytical treatment;
however, \citep{angus2014} has provided a complete description of
the amplitude scaling in the inertial regime from simulations, highlighting
the importance of not making the Boussinesq approximation. We use
two dimensional simulations of isolated filaments both to identify
the qualitative features that go into the calculation, which is especially
important for the inertial regime as it is strongly non-linear, and
also to validate comprehensively the scaling laws obtained.

\section{Velocity scaling}

\subsection{Physical picture\label{sub:Physical-picture}}

An isolated filament is, under the assumptions here, a density fluctuation
comparable to or larger than the background in amplitude, extended
along the magnetic field and with a monopolar structure perpendicular
to it. The basic mechanism of filament motion is that there is a current
source due to the inhomogeneity of the magnetic field which propels
the filament, balanced in the simple model considered here by inertia
of the surrounding background plasma and by dissipation from currents
through the sheath where the magnetic field intersects the wall at
the targets. This description is in some respects inspired by the
`equivalent circuit' picture\citep{russell2004,krasheninnikov2008}.
In contrast to the qualitative description in \citep{krasheninnikov2008},
here we consider drifts/currents only in the fluid model, rather than
considering the particle drifts. In order to obtain simple expressions
for the filament velocity, we reduce the problem to two dimensions
by assuming zero variation parallel to the magnetic field and look
only for a maximum velocity of the filament. First we sketch the physical
picture, before turning to the full calculation in Section \ref{sec:Velocity-scaling-calculation}.

\paragraph{Drive}

\begin{figure}[h]
\includegraphics[bb=0bp 670bp 390bp 810bp,clip,width=0.5\textwidth]{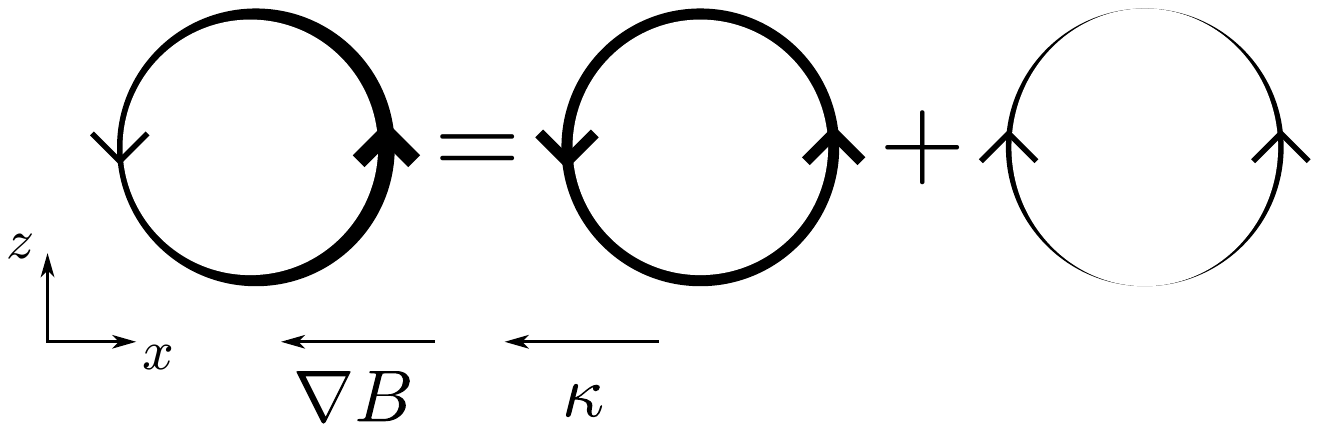}

\caption{Diamagnetic current in a filament: The width of the line represents
the magnitude of the current (integrated in the $\hat{\boldsymbol{b}}$-direction).
The diamagnetic current (left) can be split into a divergence-free
part (centre) and the remainder (right). This last, divergent part
must be closed by parallel or polarisation currents and so provides
the drive for filament motion\label{fig:diamagnetic-divergence}}
\end{figure}

Due to the curvature of the magnetic field, represented by $\boldsymbol{\kappa}=-\hat{\boldsymbol{x}}/R_{C}$
where $R_{C}$ is the radius of curvature, and the change in its magnitude
with radius, $\nabla B=-B\hat{\boldsymbol{x}}/R_{C}$, the diamagnetic
current $\boldsymbol{J}_{\mathrm{dia}}=B^{-1}\nabla(nT)\times\hat{\boldsymbol{b}}$
 ($\hat{\boldsymbol{b}}\equiv\boldsymbol{B}/B$ being the unit vector
in the direction of the magnetic field) has a non-zero divergence,
as sketched in Figure \ref{fig:diamagnetic-divergence}. This divergence
gives a current source whose strength is proportional to the pressure
gradient in the direction perpendicular to $\boldsymbol{\kappa}$
and $\nabla B$, which we will call $\hat{\boldsymbol{z}}$. In order
to maintain quasineutrality, this current must be closed. Depending
on the parameters of the filament, this may occur either through the
polarisation current (leading to inertial evolution) or the parallel
current (giving rise to sheath current dissipation). 

\paragraph{}

\paragraph{Inertial evolution}

In the absence of dissipation mechanisms (viscosity or sheath currents),
there can be no steady state, since the energy of the system increases
monotonically\citep{angus2014}. Nevertheless, the filament reaches
a maximum velocity, even without dissipation. This contrasts with
hydrodynamic drag on a solid object, where the velocity is ultimately
limited by viscous dissipation either in a surface layer around the
object or in the wake behind it. The mechanism causing the filament
velocity to saturate is that it expands as it moves, drawing in fluid
from the `background' flow. This is rather like a rocket engine
in reverse; instead of generating thrust by expelling material, the
filament resists acceleration by absorbing material. The maximum velocity
occurs when the rate at which momentum is acquired through this absorption
balances the $\boldsymbol{J}\times\boldsymbol{B}$ force from the
polarisation current.

\paragraph{Sheath current dissipation}

\begin{figure}[h]
\includegraphics[bb=20bp 650bp 410bp 800bp,clip,width=0.5\textwidth]{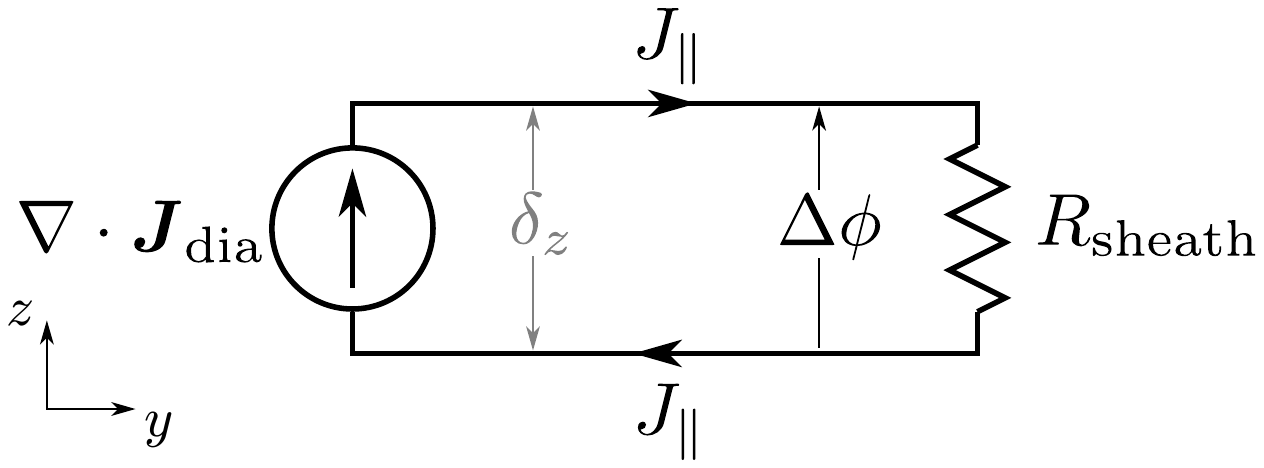}

\caption{Equivalent circuit for a filament in which the diamagnetic current
closes through the parallel current path\label{fig:sheath-circuit}}
\end{figure}

When the current source closes through the parallel current it is
regulated by the sheath boundary condition, (\ref{eq:sheath-boundary-condition}),
which determines the current that passes through the Debye sheath
to the target plates. For small currents the sheath behaves as an
ideal resistor, with current proportional to the electrostatic potential,
$\phi$. In steady state the potential is then proportional to the
current source, behaving as a simple electrical circuit (Figure \ref{fig:sheath-circuit}).
The motion of the filament is then just the $\boldsymbol{E}\times\boldsymbol{B}$
drift due to this potential.

\subsection{Scaling calculation\label{sec:Velocity-scaling-calculation}}

\begin{figure}
\includegraphics[bb=40bp 570bp 430bp 810bp,clip,width=0.5\textwidth]{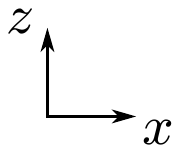}

\caption{Density (colour-map) and potential (contours; solid for $\phi>0$,
dashed for $\phi<0$) for a filament in the sheath regime, at maximum
$V_{x}$\label{fig:density-potential-configuration}}
\end{figure}

We use a coordinate system that takes $\hat{\boldsymbol{x}}$ in the
`radial' direction (anti-parallel to the magnetic field curvature
and $\nabla B$), $\hat{\boldsymbol{y}}$ along the magnetic field
and $\hat{\boldsymbol{z}}$ in the binormal direction. We take the
filament to be a monopolar density structure (possibly distorted,
in the inertial regime), characterised by two length scales, $\delta_{x}$
and $\delta_{z}$ in the $\hat{\boldsymbol{x}}$- and $\hat{\boldsymbol{z}}$-directions,
and an induced, dipolar potential structure. To aid visualisation
of the configuration, the density and potential from a simulation
(see Section \ref{sec:Comparison-with-simulations}) of a filament
in the sheath current regime are shown in Figure \ref{fig:density-potential-configuration}.
We assume cold ions and isothermal electrons, neglect electron inertia
and use Bohm normalisation (as in \citep{easy2014}, times are normalised
to the ion cyclotron frequency $\Omega_{\mathrm{ci}}=eB_{0}/m_{\mathrm{i}}$,
lengths to the hybrid Larmor radius $\rho_{\mathrm{s}}=c_{\mathrm{s}}/\Omega_{\mathrm{ci}}$,
where $c_{\mathrm{s}}=\sqrt{T_{\mathrm{e}}/m_{\mathrm{i}}}$ is the
sound speed, and electrostatic potential by $T_{\mathrm{e}}/e$, with
$e$ the elementary charge, $B_{0}$ the magnetic field strength,
$m_{\mathrm{i}}$ the ion mass and $T_{e}$ the electron temperature;
in addition densities are normalised to a typical value, usually the
background density).

The system is driven by the diamagnetic current, and its behaviour
determined by whether the divergence of the diamagnetic current closes
principally through the polarisation or the parallel current. We now
consider each current in turn.

\paragraph{Diamagnetic current}

For cold ions, the diamagnetic current is due just to the electron
diamagnetic velocity. Neglecting electron inertia and viscosity, the
electron momentum equation is
\begin{equation}
0=-\nabla n+n\nabla\phi+n\frac{\boldsymbol{V}_{\mathrm{e}}\times\boldsymbol{B}}{B_{0}}.
\end{equation}
Taking the cross product with $\hat{\boldsymbol{b}}$ gives the perpendicular
electron velocity with two contributions, the $\boldsymbol{E}\times\boldsymbol{B}$
and diamagnetic velocities: $\boldsymbol{V}_{\mathrm{e}}=\boldsymbol{V}_{E\times B}+\boldsymbol{V}_{\mathrm{dia}}$,
with
\begin{equation}
n\boldsymbol{V}_{\mathrm{dia}}=-B_{0}\frac{\nabla n\times\hat{\boldsymbol{b}}}{B}.
\end{equation}
Here we must not yet take $B=B_{0}$ ($B_{0}$ being the constant
magnetic field used for normalisation) until we have taken its gradient
in the divergence of the diamagnetic current:
\begin{eqnarray}
\nabla\cdot\boldsymbol{J}_{\mathrm{dia}} & =B_{0}\nabla\cdot\left(\frac{\nabla n\times\hat{\boldsymbol{b}}}{B}\right)\nonumber \\
 & =-B_{0}\frac{\nabla n\cdot\nabla\times\hat{\boldsymbol{b}}}{B}-B_{0}\frac{\nabla B\cdot\nabla n\times\hat{\boldsymbol{b}}}{B^{2}}\nonumber \\
 & =2\frac{\hat{\boldsymbol{z}}\cdot\nabla n}{R_{\mathrm{C}}}=-\hat{\boldsymbol{b}}\cdot\boldsymbol{g}\times\nabla n\nonumber \\
 & \sim-\frac{g}{\delta_{z}}\left(n-n_{0}\right),\label{eq:divergence_J_dia}
\end{eqnarray}
for a toroidal magnetic field $\boldsymbol{B}=\boldsymbol{e}_{\phi}/R$
at a major radius $R=R_{\mathrm{C}}$ ($R_{\mathrm{C}}$ denoting
the radius of curvature of the magnetic field and $\boldsymbol{e}_{\phi}$
being the unit vector in the toroidal direction) much larger than
the size of a filament and defining $\boldsymbol{g}=g\hat{\boldsymbol{x}}=\hat{\boldsymbol{x}}/R_{\mathrm{C}}$.

\paragraph{Polarisation current}

\begin{figure}
\includegraphics[bb=40bp 570bp 430bp 810bp,clip,width=0.5\textwidth]{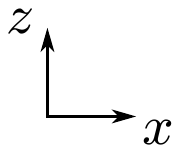}

\caption{Density (colour-map) and streamlines of flow velocity in the frame
of the filament (white) at the time when maximum velocity is reached\label{fig:n_streamlines}}
\end{figure}
\begin{figure}
\includegraphics[bb=40bp 570bp 430bp 810bp,clip,width=0.5\textwidth]{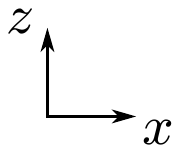}

\caption{Vorticity (colour-map) and streamlines of flow velocity in the frame
of the filament (white) at the time when maximum velocity is reached\label{fig:vort_streamlines}}
\end{figure}
\begin{figure*}
\includegraphics[width=0.5\textwidth]{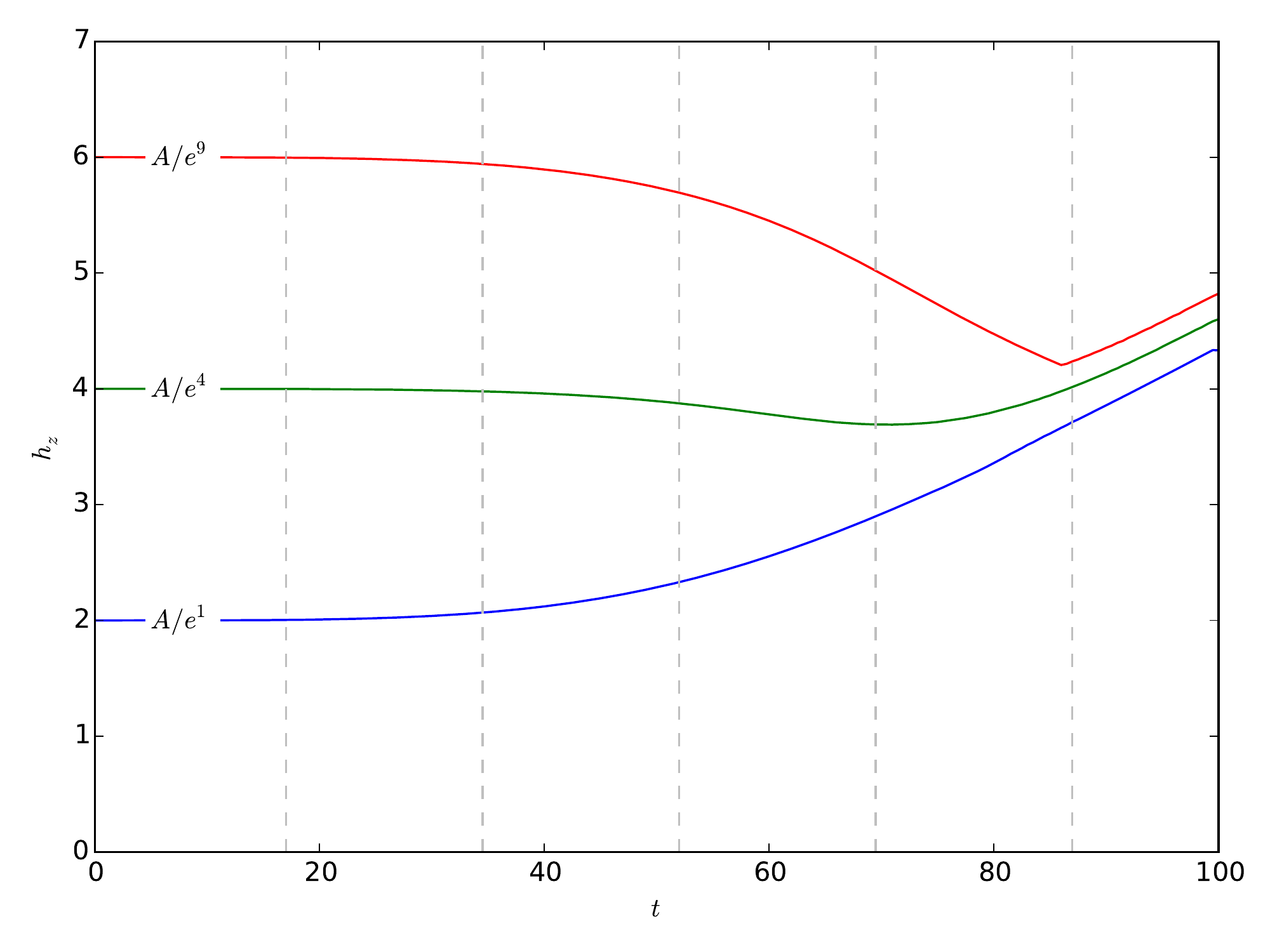}\hfill{}\includegraphics[width=0.5\textwidth]{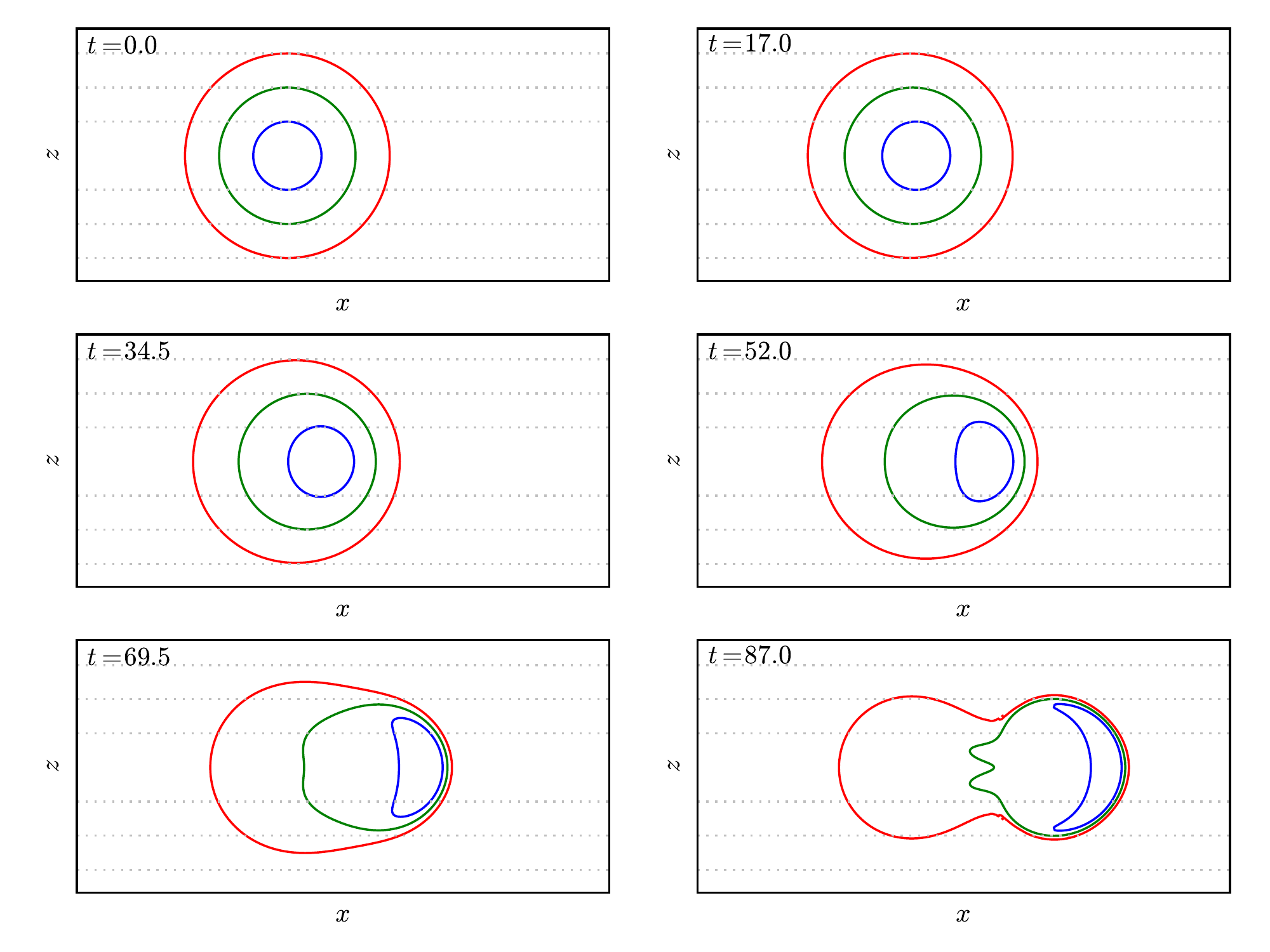}

\caption{Evolution of three density contours with time. $h_{z}$ is the height
of the contour: the difference between the maximum and minimum of
$z$ on the contour. The snapshots on the right show the contours
(at levels $n=n_{0}\left(1+A/e^{i}\right)$ for $i=1,4,9$ from inside
to outside, corresponding to the lines in the left plot) at the times
indicated by the dashed vertical lines. $t=87$ is the time at which
the filament reaches maximum velocity. The horizontal dotted lines
indicate the upper and lower extents of the contours at $t=0$\label{fig:contour_heights}}
\end{figure*}
\begin{figure}
\includegraphics[bb=40bp 570bp 430bp 810bp,clip,width=0.5\textwidth]{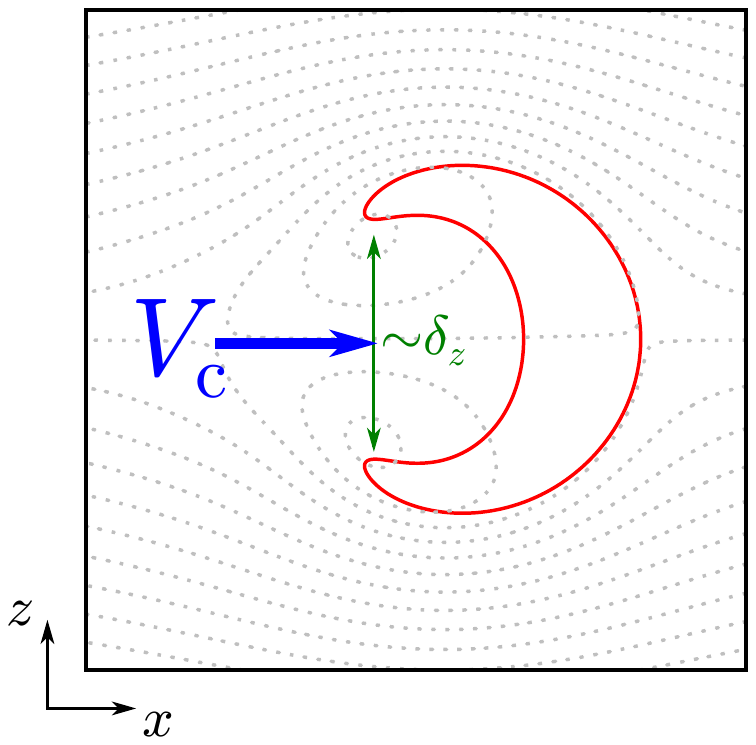}

\caption{The stationary points of the flow (the dotted lines are streamlines
of flow velocity in the frame of the filament) pin the ends of density
contours which are `inflated' by the flow between the stationary
points, with velocity $\sim V_{\mathrm{c}}$\label{fig:Vc}}
\end{figure}

Define the filament velocity, $\boldsymbol{V}_{\mathrm{f}}$, as the
velocity of the centre of mass of the perturbation,
\begin{equation}
\boldsymbol{V}_{\mathrm{f}}=\int d^{2}x\,(n-n_{0})\boldsymbol{V}
\end{equation}
and the `filament frame' as that moving at $\boldsymbol{V}_{\mathrm{f}}$
with respect to the lab frame. In the filament frame there are two
distinct regions (see figure \ref{fig:n_streamlines}%
\footnote{The figures \ref{fig:n_streamlines}, \ref{fig:vort_streamlines},
\ref{fig:contour_heights} and \ref{fig:Vc} in this section are taken
from a simulation with filament amplitude eight times the background
using the equations (\ref{eq:n-2d}), (\ref{eq:Omega-2d}) and (\ref{eq:phi-2d})
but, in contrast to Section \ref{sec:Comparison-with-simulations},
the curvature terms (proportional to $\boldsymbol{g}$) in (\ref{eq:n-2d})
are dropped and sheath currents and viscosity are dropped from (\ref{eq:Omega-2d})
(i.e. $L_{\|}\rightarrow\infty$ and $\mu_{\mathrm{i}}=0$). This
ensures that we are examining here a purely inertial filament without
$\boldsymbol{E}\times\boldsymbol{B}$ compressibility, which we do
in order to understand the qualitative features of this regime.%
}): the `exterior region' outside the filament where the velocity
is always to the left (the streamlines are open); and the `filament
region' where the velocity circulates (the streamlines are closed).

There is always a force on the filament from the polarisation current
$\boldsymbol{J}_{\mathrm{pol}}\times\boldsymbol{b}$ force, which
tends to increase its momentum (in the $x$-direction). For the velocity
to be constant this gain in momentum must be balanced. In the frame
of the background plasma, the balance is provided by absorbing stationary
background plasma, which increases the mass of the filament (without
adding extra momentum), allowing the velocity to remain constant.
Alternatively, considered in the frame of the filament, the background
plasma has negative $x$-momentum relative to the filament, providing
a loss of momentum to balance the gain due to the $\boldsymbol{J}_{\mathrm{pol}}\times\boldsymbol{b}$
force.

We need to evaluate the rate at which the filament gains momentum
by absorbing plasma from the background. $\boldsymbol{E}\times\boldsymbol{B}$
flow is (approximately) incompressible, so areas that are advected
in the flow, in particular the areas of density contours, are preserved.
Therefore an increase in the area of the filament can only be due
to its absorbing plasma from the background (due to changes in the
flow pattern so that the closed streamline region, the `filament
region,' grows).

Even filaments initialised with a Gaussian density profile (as in
the simulations described below) generate during their evolution a
much sharper boundary between `filament' and background than was
present in the initial (Gaussian) density profile, by the time they
reach maximum velocity. As we see in Figure \ref{fig:contour_heights},
the outlying parts of the initial density distribution are either
gathered into the filament, or are left behind altogether (although
only a negligibly small fraction of the total density perturbation
is left behind). The result is that by the time the filament reaches
its maximum velocity the heights of the contours of a wide range of
densities (several orders of magnitude) are gathered together close
to a single value, just \emph{inside} the closed streamline marking
the boundary of the filament (see Figure \ref{fig:n_streamlines}),
and the heights are increasing at the same rate. Therefore the plasma
absorbed from just outside the filament boundary has a density (almost)
identical to the background density, $n_{0}$.

We denote the stream function in the filament frame by $\psi$ (whose
contours are the streamlines plotted in Figures \ref{fig:n_streamlines},
\ref{fig:vort_streamlines} and \ref{fig:Vc}). $\psi$ is related
to the electrostatic potential (which is the stream function in the
lab frame) by
\begin{equation}
\psi=\phi-V_{\mathrm{f}}z.
\end{equation}
 In the exterior region the density gradients and vorticity are negligible,
as we see in figures \ref{fig:n_streamlines} and \ref{fig:vort_streamlines},
which means that $\nabla_{\perp}^{2}\phi=0$ and hence $\nabla_{\perp}^{2}\psi=0$.
 The boundary conditions are constant $\psi$ on the filament boundary
(since we define the boundary by a streamline) and $\boldsymbol{V}=-\boldsymbol{V}_{\mathrm{f}}$
asymptotically at large distances from the filament. The flow in the
exterior region is thus the solution to a linear problem and scales
linearly with the magnitude $V_{\mathrm{f}}$. The average speed of
the plasma absorbed by the filament from the external flow therefore
scales like $V_{\mathrm{f}}$.

Putting these two results together, the plasma captured as the filament
increases its area will have an average $x$-momentum density
\[
\left\langle p_{x}\right\rangle \sim-n_{0}V_{\mathrm{f}}
\]
up to some geometrical factor depending on the shape of the filament,
which we assume does not change much, the assumption being justified
\textit{a posteriori} by the fact that the result obtained describes
the simulations well.

The rate at which the filament gains momentum from the background
is then
\[
\frac{\partial P_{x}}{\partial t}=\left\langle p_{x}\right\rangle L_{\|}\frac{\partial\mathcal{A}}{\partial t}=-n_{0}V_{\mathrm{f}}L_{\|}\frac{\partial\mathcal{A}}{\partial t}
\]
where $\mathcal{A}$ is the area of the filament (the area enclosed
by the largest closed streamline). Finally we need to estimate $\partial\mathcal{A}/\partial t$.
The filament increases its area because the crescent-shaped density
contours are `inflated' (since they cannot be compressed, their
areas being preserved) by the flow on the axis, between the stagnation
points of the circulating flow within the filament, which are separated
by a distance $\sim\delta_{z}$, as illustrated in Figure \ref{fig:Vc}.
If the average velocity (in the filament frame) of that flow is $V_{\mathrm{c}}$
then 
\begin{equation}
\frac{\partial\mathcal{A}}{\partial t}\sim V_{\mathrm{c}}\delta_{z}.\label{eq:dAdt}
\end{equation}
The only velocity scale in the system is $V_{\mathrm{f}}$, so (in
the absence of any more exact calculation), we estimate $V_{\mathrm{c}}\sim V_{\mathrm{f}}$
so that
\begin{equation}
\frac{\partial P_{x}}{\partial t}\sim-L_{\|}\delta_{z}n_{0}V_{\mathrm{f}}^{2}.\label{eq:dPdt}
\end{equation}

The total force on the filament is in the positive $x$-direction,
with magnitude ($\mathcal{F}$ denotes the filament region and $\partial\mathcal{F}$
its boundary)
\begin{eqnarray}
 & -L_{\|}\int_{\mathcal{F}}d^{2}x\, J_{\mathrm{pol,}z}\nonumber \\
 & =-L_{\|}\int_{\partial\mathcal{F}}ds\, z\hat{\boldsymbol{n}}\cdot\boldsymbol{J}_{\mathrm{pol}}+L_{\|}\int_{\mathcal{F}}d^{2}x\, z\nabla\cdot\boldsymbol{J}_{\mathrm{pol}}\nonumber \\
 & =-L_{\|}\int_{\partial\mathcal{F}}ds\, z\hat{\boldsymbol{n}}\cdot\boldsymbol{J}_{\mathrm{pol}}-L_{\|}\int_{\mathcal{F}}d^{2}x\, z\nabla\cdot\boldsymbol{J}_{\mathrm{dia}}\nonumber \\
 & =-L_{\|}\int_{\partial\mathcal{F}}ds\, z\hat{\boldsymbol{n}}\cdot\boldsymbol{J}_{\mathrm{pol}}-L_{\|}\int_{\mathcal{F}}d^{2}x\, zg\frac{\partial n}{\partial z}\nonumber \\
 & =-L_{\|}\int_{\partial\mathcal{F}}ds\, z\hat{\boldsymbol{n}}\cdot\boldsymbol{J}_{\mathrm{pol}}-L_{\|}\int dx\left[zg\left(n-n_{0}\right)\right]_{z_{-}}^{z_{+}}+L_{\|}\int_{\mathcal{F}}d^{2}x\, g\left(n-n_{0}\right)\nonumber \\
 & \approx gAL_{\|}n_{0}\delta_{x}\delta_{z}\label{eq:total_force}
\end{eqnarray}
integrating by parts twice. We may neglect the surface terms because
firstly $\int_{\mathcal{E}}d^{2}x\, J_{\mathrm{pol,}z}\approx0$ in
the exterior region, $\mathcal{E}$, since the flow pattern is almost
constant and so the external plasma must be close to global force
balance. Since also $\nabla\cdot\boldsymbol{J}_{\mathrm{pol}}=0$
in the exterior region
\begin{equation}
0=\int_{\mathcal{E}}d^{2}x\, J_{\mathrm{pol},z}-\int_{\mathcal{E}}d^{2}x\, z\nabla\cdot\boldsymbol{J}_{\mathrm{pol}}=-\int_{\infty}ds\, z\hat{\boldsymbol{n}}\cdot\boldsymbol{J}_{\mathrm{pol}}+\int_{\partial\mathcal{F}}ds\, z\hat{\boldsymbol{n}}\cdot\boldsymbol{J}_{\mathrm{pol}}
\end{equation}
and $\boldsymbol{J}_{\mathrm{pol}}$ vanishes at infinity ($r^{2}\boldsymbol{J}_{\mathrm{pol}}\rightarrow0$
as $r\rightarrow\infty$ since otherwise the entire background plasma
would accelerate). Secondly $(n-n_{0})\approx0$ on the boundary,
so $\int dx\left[zg\left(n-n_{0}\right)\right]_{z_{-}}^{z_{+}}\approx0$,
where $z_{+}$ and $z_{-}$ denote the upper and lower positions of
the boundary at a particular $x$. The area integral is just the total
number of particles in the perturbation, which is conserved from the
initial state (and almost all contained within $\mathcal{F}$) and
which we may estimate as
\begin{equation}
\int_{\mathcal{F}}d^{2}x\,\left(n-n_{0}\right)\sim An_{0}\delta_{x}\delta_{z}
\end{equation}
where $A$ is the amplitude of the filament relative to the background
and $\mathcal{A}\sim\delta_{x}\delta_{z}$ is the area of the filament
(we assume that the changes in $\mathcal{A}$ with time are small
compared to its magnitude). 

Therefore the maximum filament velocity is attained when the rate
of momentum gain from the exterior plasma, (\ref{eq:dPdt}), balances
the total force, (\ref{eq:total_force}), so 
\begin{equation}
V_{\mathrm{f}}\sim\sqrt{gA\delta_{x}}\label{eq:inertial-limit}
\end{equation}

\paragraph{Parallel current}

Linearising the sheath boundary condition,
\begin{equation}
\left.J_{\|}\right|_{\mathrm{sheath}}=\pm n_{\mathrm{sheath}}\left(1-e^{-\left.\phi\right|_{\mathrm{sheath}}}\right)\approx\pm n\phi\label{eq:sheath-boundary-condition}
\end{equation}
and integrating over $y$,  
\begin{equation}
\int dy\,\nabla\cdot\boldsymbol{J}_{\|}\approx2n\phi.
\end{equation}
Since the filament velocity is approximately the central $\boldsymbol{E}\times\boldsymbol{B}$
velocity of the filament (in the lab frame), we can further estimate
$V_{\mathrm{f}}\sim\phi/\delta_{z}$ and hence
\begin{equation}
\int dy\,\nabla\cdot\boldsymbol{J}_{\|}\sim n\delta_{z}V_{\mathrm{f}}.\label{eq:divergence_J_parallel}
\end{equation}
So balancing this with the diamagnetic divergence, (\ref{eq:divergence_J_dia})
(integrated over $y$), the velocity in this regime is
\begin{equation}
V_{\mathrm{f}}\sim\frac{L_{\|}g}{\delta_{z}^{2}}\frac{\left(n-n_{0}\right)}{n}\sim\frac{L_{\|}g}{\delta_{z}^{2}}\frac{A}{1+\beta A}.\label{eq:sheath-limit}
\end{equation}
The interpretation of the factor $\left(n-n_{0}\right)/n\sim A/\left(1+\beta A\right)$
and the value of the constant $\beta$ will be discussed in Section
(\ref{sub:Amplitude-scaling}).

We have two regimes for filament dynamics, the inertial regime with
maximum velocity given by (\ref{eq:inertial-limit}) for narrow filaments
and the sheath current regime with velocity given by (\ref{eq:sheath-limit})
for wide filaments. Interestingly, the velocity depends on $\delta_{x}$
but not $\delta_{z}$ in the inertial regime and conversely on $\delta_{z}$
but not $\delta_{x}$ in the sheath current regime. The latter point
was noted in the seminal paper \citep{krasheninnikov2001}, in the
case without background plasma where a separable solution exists,
but seems to have been neglected since then. The transition between
the two regimens occurs when the velocities (\ref{eq:inertial-limit})
and (\ref{eq:sheath-limit}) are comparable, so
\begin{equation}
\sqrt{\delta_{x}}\delta_{z}^{2}\sim\sqrt{g}L_{\|}\frac{\sqrt{A}}{1+\beta A}.\label{eq:inertial-sheath-cross}
\end{equation}

\section{Comparison with simulations\label{sec:Comparison-with-simulations}}

We have used two dimensional simulations to validate the scalings
derived in Section \ref{sec:Velocity-scaling-calculation}. The equations
used, based on those in \citep{easy2014}, represent a filament assumed
to have negligible variation along the magnetic field; closure is
given by integrating the three dimensional system over the parallel
direction to give equations for the density and vorticity
\begin{equation}
\frac{dn}{dt}=\hat{\boldsymbol{b}}\cdot\boldsymbol{g}\times\left(n\nabla\phi-\nabla n\right)+\frac{n\left(1-e^{-\phi}\right)}{L_{\|}}+\mu_{n}\nabla_{\perp}^{2}n\label{eq:n-2d}
\end{equation}
\begin{equation}
\frac{d\Omega}{dt}=-\frac{1}{2}\hat{\boldsymbol{b}}\cdot\nabla V_{E\times B}^{2}\times\nabla n-\hat{\boldsymbol{b}}\cdot\boldsymbol{g}\times\nabla n+\frac{n\left(1-e^{-\phi}\right)}{L_{\|}}+\mu_{\mathrm{i}}\nabla_{\perp}^{2}\Omega\label{eq:Omega-2d}
\end{equation}
\begin{equation}
\Omega=\nabla\cdot\left(n\nabla_{\perp}\phi\right)\label{eq:phi-2d}
\end{equation}
with $d/dt=\left(\partial/\partial t+\hat{\boldsymbol{b}}\cdot\nabla\phi\times\nabla\right)$,
$\boldsymbol{V}_{E\times B}=\hat{\boldsymbol{b}}\times\nabla\phi$
and, as before, $\boldsymbol{g}=g\hat{\boldsymbol{x}}$. (\ref{eq:Omega-2d})
may be derived by taking the parallel component of the curl of the
ion momentum equation
\begin{equation}
\frac{d\left(n\boldsymbol{V}_{\mathrm{i}}\right)}{dt}=\boldsymbol{J}_{\mathrm{pol}}\times\hat{\boldsymbol{b}}+\mu_{\mathrm{i}}\nabla_{\perp}^{2}\left(n\boldsymbol{V}_{\mathrm{i}}\right)
\end{equation}
and replacing $\nabla\cdot\boldsymbol{J}_{\mathrm{pol}}$ using quasineutrality,
$\nabla\cdot\boldsymbol{J}_{\mathrm{dia}}+\nabla\cdot\boldsymbol{J}_{\mathrm{pol}}+\nabla\cdot\boldsymbol{J}_{\|}=0$.
(\ref{eq:phi-2d}) is inverted for $\phi$ using a new multigrid solver
currently under development for BOUT++, to be reported elsewhere.
The dimensionless parameters used are $L_{\|}=11000$, $g=2.5\times10^{-3}$,
$\mu_{n}=1.5\times10^{-5}$ and $\mu_{\mathrm{i}}=4\times10^{-4}$;
for this theoretical study the dissipative parameters have been reduced
by a factor of 100 compared to those used in \citep{easy2014}, based
on the expressions in \citep{fundamenski2007}, in order to minimise
their effect on the inertial regime, since we do not treat viscosity
here. Filaments are initialised as Gaussian density fluctuations with
elliptical contours, tilted at an angle $\alpha$ to the $\hat{\boldsymbol{x}}$-direction,
on a constant background,
\begin{equation}
n(t=0)=n_{0}\left(1+A\exp\left(-\frac{x'^{2}/\epsilon+\epsilon z'^{2}}{\delta^{2}}\right)\right)
\end{equation}
where $A$ is the amplitude of the filament, $\epsilon$ is the ratio
of the lengths of the axes of the ellipse, $\delta$ is the geometric
mean of the lengths of the axes, $x'=x\cos\alpha+z\sin\alpha$ and
$z'=z\cos\alpha-x\sin\alpha$. 
\begin{figure}
\includegraphics[bb=40bp 630bp 430bp 770bp,clip,width=0.5\textwidth]{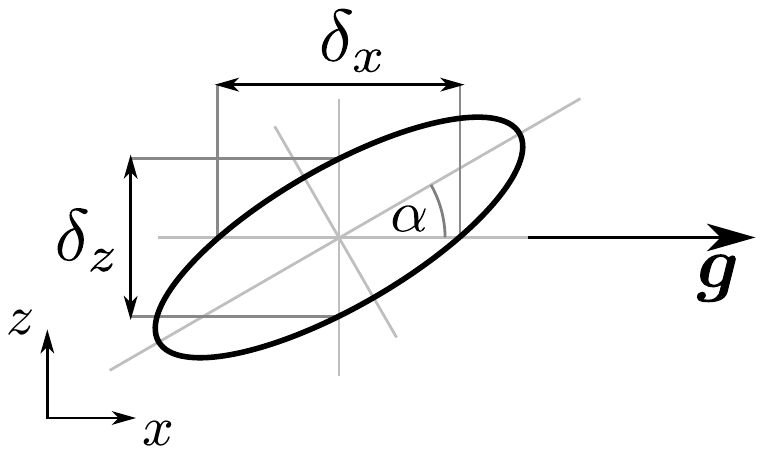}

\caption{Schematic showing sizes and orientation of a tilted, elliptical filament
cross-section\label{fig:schematic-cross-section}}
\end{figure}
The diagram in figure \ref{fig:schematic-cross-section} shows the
configuration with the corresponding length scales, in the $\hat{\boldsymbol{x}}$-
and $\hat{\boldsymbol{z}}$-directions respectively, 
\begin{eqnarray}
\delta_{x} & =\delta\sqrt{\frac{\epsilon}{\left(\cos^{2}\alpha+\epsilon^{2}\sin^{2}\alpha\right)}}\\
\delta_{z} & =\delta\sqrt{\frac{\epsilon}{\left(\epsilon^{2}\cos^{2}\alpha+\sin^{2}\alpha\right)}},
\end{eqnarray}
 which appear in the velocity scaling above. Filament velocities are
measured as the maximum velocity in the $\hat{\boldsymbol{x}}$-direction
of the centre of mass of the density above background, i.e.~$n-n_{0}$.
These simulations, as well as the three dimensional ones in Section
\ref{sec:3d}, have been implemented using BOUT++\citep{dudson2009,dudson2015}.

\subsection{Scaling with size and shape\label{sub:Scaling-with-size-and-shape}}

\begin{figure*}
\includegraphics[width=1\textwidth]{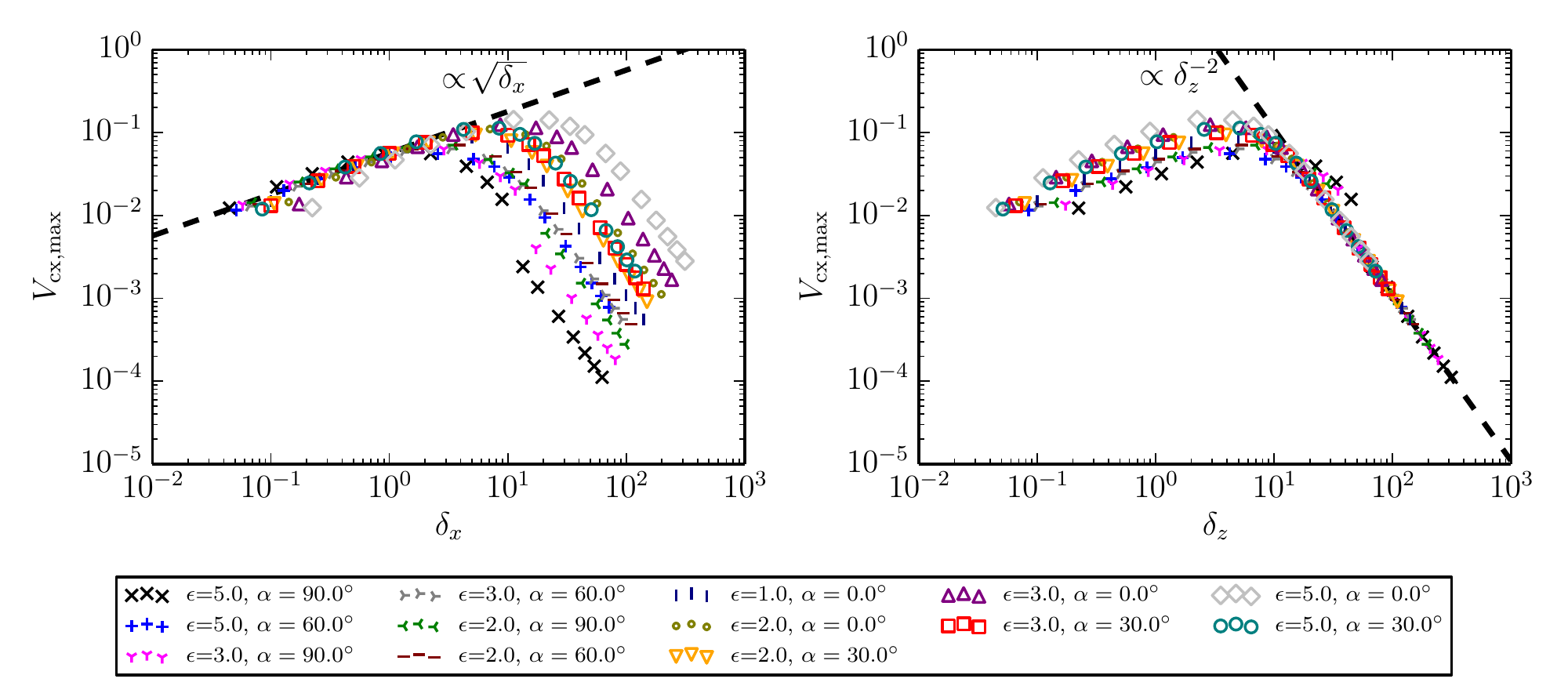}

\caption{Inertial scaling vs.~$\delta_{x}$ (left) and sheath current scaling
vs.~$\delta_{z}$ (right) with $A=4$ for several values of $\epsilon$
and $\alpha$\label{fig:parameter-scan}}
\end{figure*}

The results of a scan in width $\delta$, ellipticity $\epsilon$
and inclination $\alpha$, at constant amplitude $A=4$, are shown
in Figure \ref{fig:parameter-scan}. For four different $\epsilon$
and four different $\alpha$, scanning $\delta$ results in a region
of $\sqrt{\delta_{x}}$ scaling, corresponding to the inertial regime,
and a region of $\delta_{z}^{-2}$ scaling corresponding to the sheath
current regime. Moreover, although we have a three parameter space,
$\{\delta,\epsilon,\alpha\}$, of filament sizes and shapes, a single
combination ($\delta_{x}$ in the inertial regime and $\delta_{z}$
in the sheath current regime) entirely determines the filament velocity,
at a particular amplitude (the effect of amplitude will be discussed
in Section \ref{sub:Amplitude-scaling}); this is evident from Figure
\ref{fig:parameter-scan} since all the points, for any $\epsilon$
and $\alpha$ (in the region where they follow $\sqrt{\delta_{x}}$
or $\delta_{z}^{-2}$ scaling) have not only the same gradient but
also the same absolute magnitude.

The results for $\epsilon=3$, $\alpha=90^{\circ}$ and $\epsilon=5$,
$\alpha=90^{\circ}$ exceed the $\delta_{z}^{-2}$ scaling for $\delta_{z}\approx20-50$.
This is due to the non-linearity of the sheath boundary condition
which is included in the simulations, but not in the scaling. The
maximum value of the potential is $\phi\approx1.0$ at $\delta_{z}\approx26$
for $\epsilon=3$ and $\phi\approx1.0$ at $\delta_{z}\approx34$
for $\epsilon=5$, so it makes sense that the deviation of $\exp\left(-\phi\right)$
from its linearised form is noticeable here. The effect of the non-linearity
is to decrease the magnitude of the negative lobe of potential and
increase the positive lobe. However, the enhancement of the positive
lobe must be larger in order to allow the same magnitude, $j_{0}$,
of current through the sheath: $\left|\phi_{+}\right|/\left|\phi_{-}\right|=-\ln\left(1-j_{0}\right)/\ln\left(1+j_{0}\right)>1$.
The overall effect is therefore to increase the filament velocity,
since the increase from the positive lobe outweighs the decrease from
the negative lobe, resulting in a larger $E_{z}$.

The trend lines are $V\approx5.7\times10^{-2}\sqrt{\delta_{x}}$ for
the inertial scaling and $V\approx11\delta_{z}^{-2}$ for the sheath
scaling. These intersect at $\delta_{z}^{4/5}\delta_{x}^{1/5}\approx8.2$
whereas (\ref{eq:inertial-sheath-cross}) suggests a crossover at
$\delta_{z}^{4/5}\delta_{x}^{1/5}\approx11.9$ for these parameters
(using $\beta$ from Section (\ref{sub:Amplitude-scaling})), in good
agreement up to the order unity factors which are not fixed by analytical
scaling arguments.

\subsection{Amplitude scaling\label{sub:Amplitude-scaling}}

\begin{figure*}
\includegraphics[bb=0bp 10bp 288bp 216bp,clip,width=0.5\textwidth]{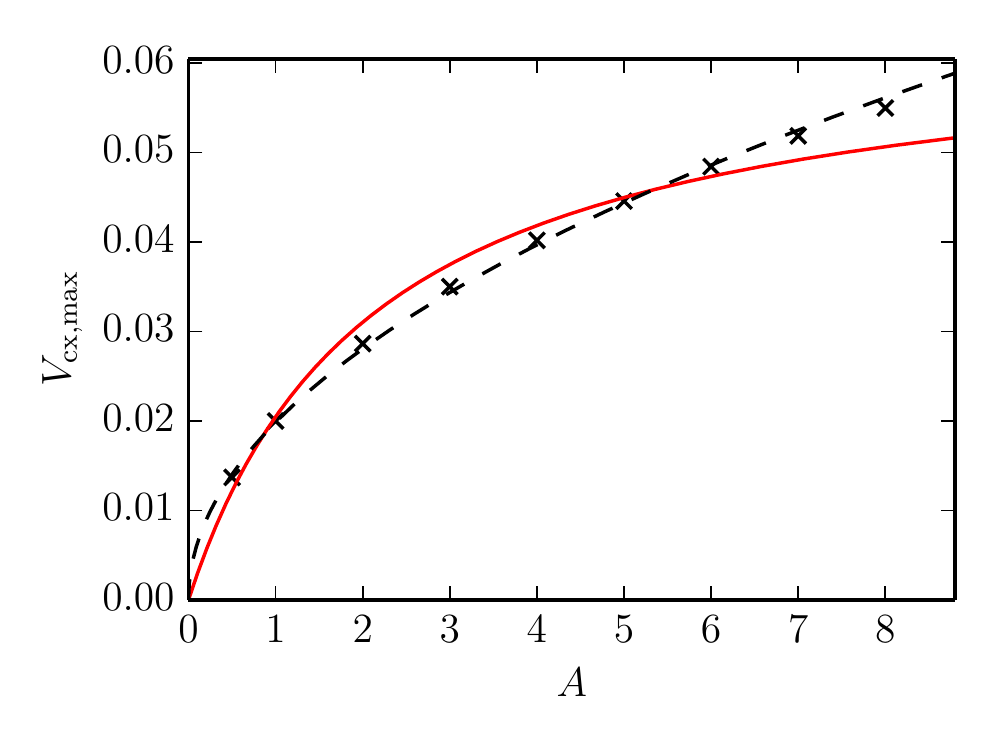}\includegraphics[bb=0bp 10bp 288bp 216bp,clip,width=0.5\textwidth]{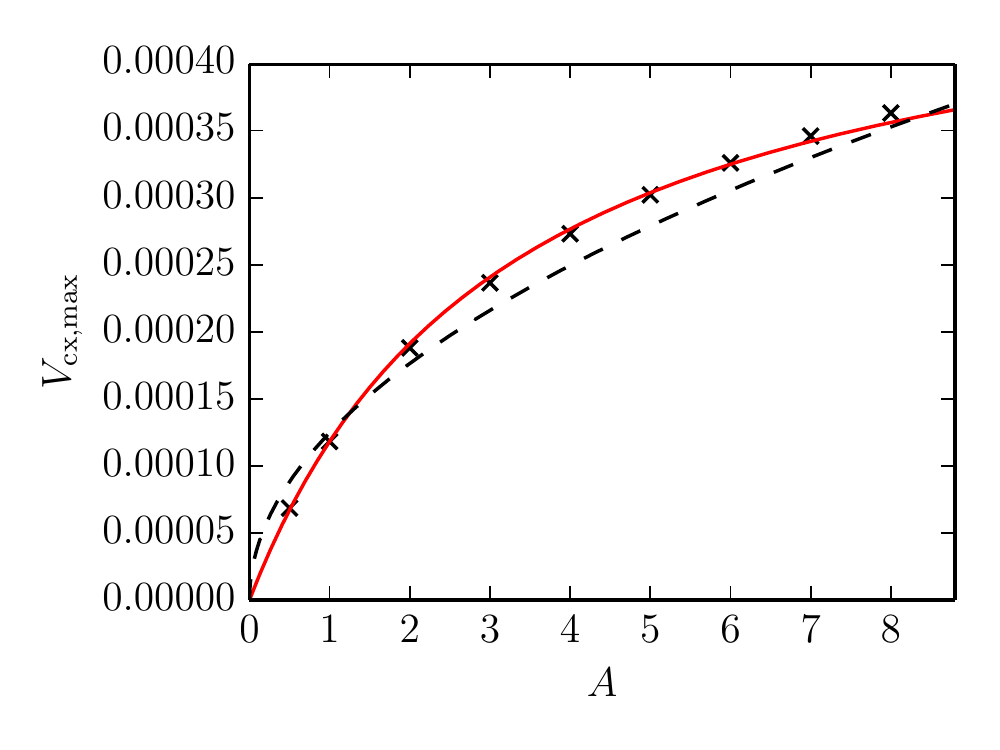}

\caption{Amplitude scaling in the inertial regime, $\delta_{x}=\delta_{z}=0.5$,
(left) and in the sheath current regime, $\delta_{x}=\delta_{z}=200$,
(right). The solid red line shows the fitted ansatz, (\ref{eq:ansatz_lin}),
and the dashed black line shows best fit line $\propto\sqrt{A}$\label{fig:amplitude-scan}}
\end{figure*}

The amplitude scaling predicted for the inertial regime is simple,
just $V\propto\sqrt{A}$, where the maximum fluctuation amplitude
is $A=\max\left[\left(n-n_{0}\right)/n_{0}\right]$. For the sheath
current regime the prediction is a little more complicated. We may
estimate $\nabla_{z}(n-n_{0})\sim An_{0}/\delta_{z}$, but we need
to find a representative value of $n$ to estimate $n^{-1}\nabla_{z}(n-n_{0})$
in (\ref{eq:sheath-limit}). We find, by fitting the simulation results,
that the appropriate value is neither the maximum, $n\sim(1+A)n_{0}$
nor the background $n\sim n_{0}$, but rather an intermediate value
$n\sim(1+\beta A)n_{0}$, where $0<\beta<1$. $\beta$ corresponds
to some point on the density profile at a distance $\tilde{\delta}_{z}$:
$\beta=\exp(-\tilde{\delta}_{z}^{2}/\delta_{z}^{2})$. Then $V_{\mathrm{f}}$
as a function of $A$ with all other parameters held constant is
\begin{equation}
V_{\mathrm{f}}\propto\frac{A}{1+\beta A}\label{eq:ansatz_lin}
\end{equation}
if the representative value of $n$ is found at the same relative
position, $\tilde{\delta}_{z}/\delta_{z}$ for any $A$. This does
indeed seem to be the case, as illustrated in Figure \ref{fig:amplitude-scan}.
For an amplitude scan with $\delta_{x}=\delta_{z}=0.5$, which is
in the inertial regime $V\propto\sqrt{A}$ fits well, in agreement
with \citep{angus2014}, with $\varepsilon\approx0.016$, where $\varepsilon$
is the root-mean-square relative error. For an amplitude scan with
$\delta_{x}=\delta_{z}=200$, which is in the sheath current regime,
(\ref{eq:ansatz_lin}) is a good fit, $\varepsilon\approx0.014$,
and gives $\beta\approx0.31$. These conclusions also hold when varying
the ellipticity, see \ref{sec:Amplitude-scaling-fits}. 

Comparing this to previous work: Angus and Krasheninnikov\citep{angus2014}
established numerically the $\sqrt{A}$ scaling in the inertial regime,
which we have tried to explain analytically above; Kube and Garcia\citep{kube2011}
consider the amplitude dependence in some detail, where the scaling
analysis starts from the vorticity equation and finds similar results
to those given here, except that they use the Boussinesq approximation
in their simulations and so do not find $\sqrt{A}$ in the inertial
regime, while in the sheath current regime they assume a form equivalent
to setting $\beta=1$ which prevents them from finding a quantitative,
analytical amplitude scaling (their fit coefficients vary with amplitude,
whereas here we have only a single, constant coefficient, $\beta$,
to be derived from simulations); Angus et al.\citep{angus2012} give,
albeit briefly, a derivation starting from the momentum equation,
but neglect the amplitude dependence of the drive; Theiler et al.\citep{theiler2009}
in contrast use an interchange instability growth rate to estimate
$\partial/\partial t\sim\gamma_{\mathrm{interchange}}$, and find
a linear scaling with amplitude of the filament velocity in the inertial
regime (which is contradicted here, emphasising that it is the non-linear,
advective, term that is relevant); their derivation in \citep{theiler2009}
follows Garcia et al.\citep{garcia2005,garcia2006}, but there the
`ideal interchange rate' is defined, without explanation, as
\[
\gamma=\left(\frac{g}{\ell}\frac{\Delta\theta}{\Theta}\right)^{1/2}
\]
(where $g$ corresponds to our $g$, $\ell$ to $L_{\|}$, $\Delta\theta$
is the fluctuation amplitude and $\Theta$ is the background density)
which includes an amplitude dependence giving a square-root scaling
of the filament velocity in the inertial regime.

\section{Three Dimensional validation\label{sec:3d}}

\begin{figure}
\includegraphics[width=0.5\textwidth]{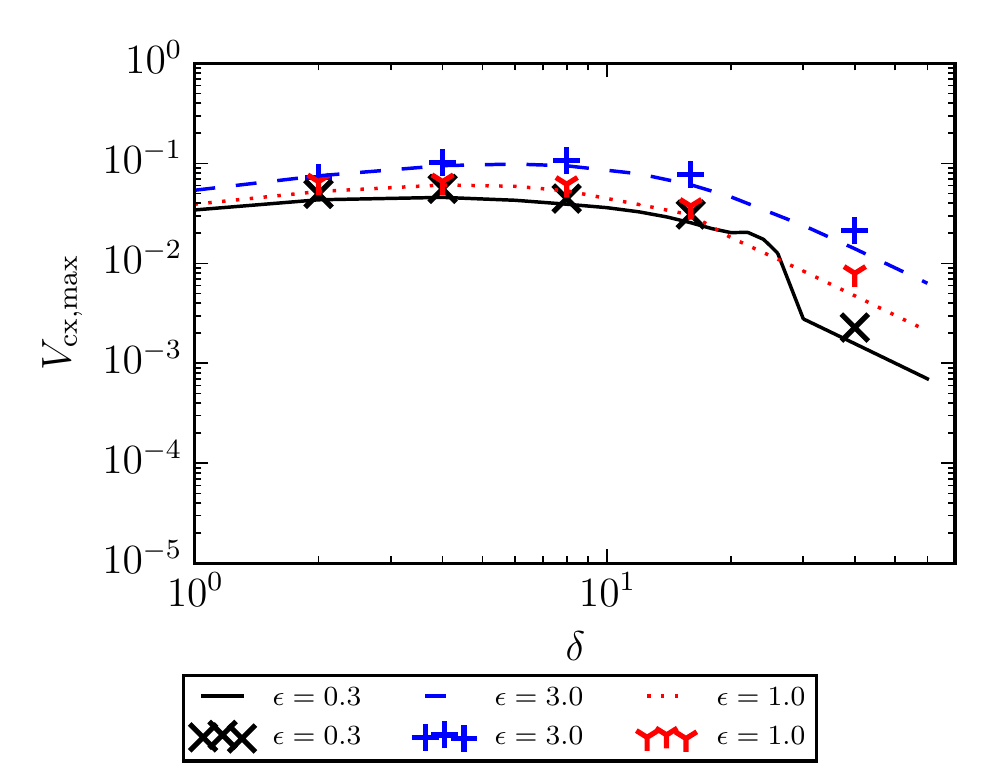}

\caption{Comparison of two dimensional (lines) and three dimensional (markers)
results for $A=2$ and several $\epsilon$\label{fig:3d-comparison}}
\end{figure}
For reasons of computational expense, the three dimensional simulations
in this section were done using the Boussinesq approximation (as were
the two dimensional simulations just in this section, for consistency).
This approximation is known to give the wrong amplitude scaling\citep{angus2014},
but seems not to affect much the scaling with size and shape (at least
the $\sqrt{\delta_{x}}$ inertial regime and $\delta_{z}^{-2}$ sheath
current regime may be found either with or without the approximation).
Here the normalised background density in the two dimensional simulations
was set to $n_{0}\approx1.48$ to be consistent with the (source-driven)
background used for the three dimensional simulations, which normalise
to the equilibrium density at the sheath entrance. The amplitude was
set to a lower value than in the two dimensional case, $A=2$, to
avoid the possibility of drift wave instabilities playing a role in
the three dimensional simulations. For more details on the equations
used in and implementation of the three dimensional simulations, see
\citep{easy2014}. The three dimensional filaments were initialised
with no variation in the parallel direction, in order to correspond
to the two dimensional calculation as closely as possible; investigation
of really `three dimensional' effects due to parallel gradients
is left for future work. Figure \ref{fig:3d-comparison} shows that
the three dimensional simulations follow the two dimensional trends
closely.

The velocity of three dimensional filaments in the sheath current
regime is slightly larger than the corresponding two dimensional ones.
We attribute this to the variation in the background density: since
this decreases near the sheath, the potential needed to drive the
same sheath current is slightly increased and it follows that the
filament velocity must then increase slightly.

The bump on the $\epsilon=1/3$ curve between $\delta=20$ and $\delta=30$,
the part most affected by the non-linearity of the boundary condition
(see Section \ref{sub:Scaling-with-size-and-shape}), is due to the
eventual fragmentation and subsequent acceleration of the filaments.
In other words, it is due to their ceasing to travel as coherent structures
and so its investigation is a subject beyond the scope of this paper.

\section{Discussion}

A number of works\citep{myra2005,theiler2009} have analysed filament
motion by analogy with the theory of linear instabilities; this has
been termed the `blob correspondence principle'\citep{myra2005,myra2006}.
Our analysis above shows that the picture is not quite so simple;
the nature of the filaments as coherent, non-linear objects is important.
Even in the sheath current regime, the density and potential fields
have qualitatively different structures, being respectively monopolar
and dipolar (Figure \ref{fig:density-potential-configuration}), rather
than the identical structures, up to a phase shift, that they would
have in linear theory; in the inertial regime the difference in the
form of the fields from a simple mode structure is even more marked
(Figures \ref{fig:n_streamlines} and \ref{fig:vort_streamlines}).
 Thus the nature of the physical processes involved in limiting filament
velocity is clearer when considering the filament itself instead of
analysing the governing equations by analogy or `correspondence'
to linear theory.

It is notable that in the inertial regime rather fine structures are
formed, especially in the vorticity (see Figure \ref{fig:vort_streamlines}).
This suggests that both finite Larmor radius effects and viscosity
may have significant effects on filaments in the inertial regime,
which could provide an interesting topic for future study, perhaps
using a gyrofluid model.

\section{Conclusions}

We have given here the first calculation of SOL filament velocity
to include the effect of the filament shape and have also clarified
the role of the filament amplitude. The analytical scaling calculations
have been extensively validated by two dimensional simulation results,
and also compared (with good agreement) to three dimensional simulations
in which the filaments are initialised without parallel variation.
Thus we now have a complete understanding of the mechanisms of filament
propagation in the simple limit considered here. This understanding
provides a solid foundation for the interpretation of filament motion
in more complicated, more realistic models.

\section*{Acknowledgements}

We would like to thank Prof.~Steve Cowley for stimulating discussions
which provided the initial impetus for the work described here. We
are grateful to one of the anonymous referees of this paper for pointing
out the critical importance of not making the Boussinesq approximation
to get the correct amplitude scaling. We are indebted to Kab Seok
Kang for his very timely work on the multigrid solver code. This work
has been carried out within the framework of the EUROfusion Consortium
and has received funding from the Euratom research and training programme
2014-2018 under grant agreement No 633053 and from the RCUK Energy
Programme {[}grant number EP/I501045{]}. To obtain further information
on the data and models underlying this paper please contact PublicationsManager@ccfe.ac.uk.
The views and opinions expressed herein do not necessarily reflect
those of the European Commission. This work used the ARCHER UK National
Supercomputing Service (http://www.archer.ac.uk) under the Plasma
HEC Consortium EPSRC grant number EP/L000237/1.

\appendix

\section{Amplitude scaling fits\label{sec:Amplitude-scaling-fits}}

In order to verify that the form of the amplitude scaling is independent
of the shape of the filament, we repeated the scans described in Section
\ref{sub:Amplitude-scaling} for several different ellipticities,
keeping $\delta_{x}$ constant for the filaments in the inertial regime
and $\delta_{z}$ constant for the filaments in the sheath current
regime so that (according to (\ref{eq:inertial-limit}) and (\ref{eq:sheath-limit}))
the velocities of the filaments should be independent of the ellipticity.

The root-mean-square relative error, $\varepsilon$, is defined by
$\varepsilon^{2}=\frac{1}{N}\sum_{i}\frac{\left(V_{i}-V(A_{i})\right)^{2}}{V_{i}^{2}}$,
where $V_{i}$ is the velocity measured from the simulation with amplitude
$A_{i}$, $V(A)$ is the predicted scaling and $N$ is the number
of simulations in the scan.

\begin{table}[H]
\begin{tabular}{|c|c|}
\hline 
$\delta_{z}$ & $\varepsilon$\tabularnewline
\hline 
\hline 
$0.5/5$ & 0.052473\tabularnewline
\hline 
$0.5/3$ & 0.039863\tabularnewline
\hline 
$0.5/2$ & 0.029693\tabularnewline
\hline 
$0.5$ & 0.015660\tabularnewline
\hline 
$0.5\times2$ & 0.022420\tabularnewline
\hline 
$0.5\times3$ & 0.031436\tabularnewline
\hline 
$0.5\times5$ & 0.044025\tabularnewline
\hline 
\end{tabular}

\caption{Inertial regime: Comparison of $\varepsilon$ for the scaling $V\propto\sqrt{A}$
at $\delta_{x}=0.5$ for several values of $\delta_{z}$. Amplitudes
used for the analysis were $A=1,1.5,2,3,4,5,6,7,8$.\label{tab:inertial}}
\end{table}
\begin{table}[H]
\begin{tabular}{|c|c|c|}
\hline 
$\delta_{x}$ & $\beta$ & $\varepsilon$\tabularnewline
\hline 
\hline 
$200/5$ & 0.30985 & 0.014128\tabularnewline
\hline 
$200/3$ & 0.30994 & 0.014123\tabularnewline
\hline 
$200/2$ & 0.30997 & 0.014123\tabularnewline
\hline 
$200$ & 0.30999 & 0.014121\tabularnewline
\hline 
$200\times2$ & 0.30999 & 0.014120\tabularnewline
\hline 
$200\times3$ & 0.30999 & 0.014121\tabularnewline
\hline 
$200\times5$ & 0.31000 & 0.014122\tabularnewline
\hline 
\end{tabular}

\caption{Sheath current regime: Comparison of (i) values of $\beta$ inferred
by least squares regression on the relative error between the ansatz
(\ref{eq:ansatz_lin}) and filament velocities measured from simulations
and (ii) values of $\varepsilon$ for this regression, at $\delta_{z}=200$
for several values of $\delta_{x}$. Amplitudes used for the analysis
were $A=1,1.5,2,3,4,5,6,7,8$.\label{tab:sheath}}
\end{table}

\bibliographystyle{iopart-num}
\bibliography{references}

\end{document}